\documentclass[aps,prb,reprint,10pt,fleqn,showpacs,showkeys,floatfix,superscriptaddress]{revtex4-1}
\usepackage{graphicx}
\usepackage{amssymb}
\usepackage{amsmath}
\usepackage{bm}
\usepackage{dcolumn}

\begin{document}


\title{Structural and Magnetic Properties of CoO-Pt core-shell nanoparticles}

\author{Adriana Zele\v{n}\'{a}kov\'{a}}
\email{adriana.zelenakova@upjs.sk}
\affiliation{Department of Condensed Matter Physics, P.~J.~\v{S}af\'{a}rik University, 
Park Angelinum 9, 041 54 Ko\v{s}ice, Slovakia}
\author{Vladimir Zele\v{n}\'{a}k}
\affiliation{Department of Inorganic Chemistry, P.~J.~\v{S}af\'{a}rik University, 
Moyzesova 11, 041 54 Ko\v{s}ice, Slovakia}
\author{\v{S}tefan Michal\'{\i}k}
\affiliation{Department of Condensed Matter Physics, P.~J.~\v{S}af\'{a}rik University, 
Park Angelinum 9, 041 54 Ko\v{s}ice, Slovakia}
\affiliation{Institute of Physics, Czech Academy of Sciences, Na Slovance 2, 
182 21 Prague, Czech Republic}
\author{Jozef Kov\'{a}\u{c}}
\affiliation{Institute of Experimental Physics, Slovak Academy of Sciences, 
04101 Košice, Slovakia}
\author{Mark W.~Meisel}
\affiliation{Department of Condensed Matter Physics, P.~J.~\v{S}af\'{a}rik University, 
Park Angelinum 9, 041 54 Ko\v{s}ice, Slovakia}
\affiliation{Department of Physics and NHMFL, University of Florida, Gainesville, 
FL 32611-8440, USA}

\date{\today}

\begin{abstract}
Using microemulsion methods, CoO-Pt core-shell nanoparticles (NPs), with 
diameters of nominally 4 nm, were synthesized and characterized 
by high-resolution transmission electron microscopy (HRTEM) and a suite of 
x-ray spectroscopies, including diffraction (XRD), absorption (XAS), 
absorption near-edge structure (XANES), and extended absorption fine 
structure (EXAFS), which confirmed the existence of CoO cores and pure Pt 
surface layers. Using a commercial magnetometer, the ac and dc magnetic 
properties were investigated over a range of temperature 
(2~K $\leq T \leq$ 300~K), magnetic field ($\leq 50$~kOe), and frequency 
($\leq 1$~kHz).  The data indicate the presence of two different magnetic 
regimes whose onsets are identified by two maxima in the magnetic signals, 
with a narrow maximum centered at 6~K and a large one centered at 
37~K. The magnetic responses in these two regimes exhibit different 
frequency dependences, where the maximum at high temperature 
follows a Vogel-Fulcher law, indicating a superparamagnetic (SPM) blocking 
of interacting nanoparticle moments and the maximum at low temperature  
possesses a power law response characteristic of a collective freezing of 
the nanoparticle moments in a superspin glass (SSG) state. This 
co-existence of blocking and freezing behaviors is consistent with 
the nanoparticles possessing an antiferromagnetically ordered core, 
with an uncompensated magnetic moment, and a magnetically disordered 
interlayer between CoO core and Pt shell.
\end{abstract}

\maketitle


\section{Introduction}

Over the past few decades, cobalt based nanoparticles (NPs) have attracted a 
significant amount of research interest because of 
their applications in ferrofluids, 
electronic components, solar energy transformers, anodes for batteries, and 
chemical catalysts.\cite{Guo201211770,Zeisberger2007224,osorio-cantillo:07B324} 
Such nanoparticles are one of the leading candidates for high 
density magnetic recording media, where the particles with small size, 
narrow size distribution, and controlled shape are required.\cite{doi:10.1021/jp027759c}  

In addition, fine Co-based particles are model materials for fundamental 
investigations of a variety of magnetic phenomena, such as exchange spring 
and exchange bias in magnetically hard and soft phases of antiferromagnetic (AFM) and 
ferromagnetic (FM) interlayers, respectively.\cite{PhysRevLett.102.097201,PhysRevB.86.014426}    
Studies of the dynamics of the Co based nanoclusters showed interesting features 
such as pure superparamagnetic (SPM) relaxation,\cite{Jonsson2004131,Tadic200912,PhysRevB.64.174416} 
spin canting,\cite{PhysRevB.86.224407} and superspin glass (SSG)  
behavior.\cite{PhysRevB.86.014426,PhysRevB.65.134406} More specifically, the SSG behavior 
is typically manifested in strongly interacting and dense nanoparticle systems showing 
spin glass (SG) behavior. The evidence of SSG transition in fine-particle systems is strengthened by 
standard spin glass fingerprints, namely the critical slowing down of the relaxation and 
the divergence of the nonlinear susceptibility at a finite glass transition temperature 
$T_g$.\cite{PhysRevB.65.134406} Finally, based on numerous studies, interparticle 
dipole-dipole interactions are known to increase the average blocking temperature and 
affect the height and distribution of the energy barriers.\cite{Morup2010}  

All of these interesting magnetic phenomena in nanoparticles 
can be induced and influenced by a combination of 
particle sizes and surface effects.\cite{Zysler2003233,Thota20071951,PhysRevB.72.132409,
0957-4484-19-18-185702,PhysRevB.71.104408,Jamet2001293,PhysRevB.60.12918,PhysRevLett.76.1743,
PhysRevB.31.2851,PhysRevLett.24.1485,Golosovsky,PhysRevB.72.104433,reviewCoreShell} 
Especially in small nanoparticles, the surface effects play an important role in tuning of the 
magnetic behavior as the decreasing of particle size leads to an increasing fraction of atoms 
lying at or near the surface, where uncompensated surface spins can generate a net magnetic moment.
\cite{PhysRevB.64.174420}  These interface effects are more pronounced 
in antiferromagnetic nanoparticles than ferromagnetic ones because of lower magnetic moment of their 
cores. Anomalous magnetic properties arising from complicated surface effects have been the focus 
of experimental studes over the past few years by investigations involving antiferromagnetic nanoparticles 
of NiO,\cite{PhysRevB.72.132409,0957-4484-19-18-185702,PhysRevB.82.054417,Rall2012} 
$\alpha$-Fe$_2$O$_3$,\cite{PhysRevB.74.012411}, FeOOH$\cdot n$H$_2$O,\cite{Punnoose2005168} and CoO.\cite{Zhang2003111,PhysRevB.82.094433,Gubin2002}  In fact, CoO is specifically germane to the present study, and its bulk form exhibits 
antiferromagnetic ordering at a N\'eel temperature $T_{\mathrm{N}}=298$~K.  
Above $T_{\mathrm{N}}$, CoO possesses a NaCl structure, whereas below $T_{\mathrm{N}}$, unstressed crystals  experience tetragonal contractions along the cubic [100] directions, giving rise to 
domains.\cite{Nature1950, PhysRevLett.26.1483,PhysRevB.24.419}  
Reducing the particle size to the nanoscopic scale reduces $T_{\mathrm{N}}$ and dramatically changes the magnetic properties.\cite{PhysRevB.82.094433}

Given the broad range of work reported, it is noteworthy that investigations involving antiferromagnetic 
CoO nanoparticles with Pt coatings has not been reported.  In this paper, we describe the detailed 
structural and magnetic study of CoO-Pt core-shell nanoparticles with diameters near 4 nm, where the choices of the Pt coating and the size of particles were driven by the desire to avoid agglomeration and to enhance potential biomedical applications.   
Due to the small size of the nanoparticles being studied, the detailed structure properties 
were investigated using the methods of X-ray absorption near-edge structure (XANES) and extended X-ray 
absorption fine structure (EXAFS) giving information about the number and type of neighboring atoms, 
inter-atomic distances, and disorder. Theoretical simulations were performed to illustrate the 
effect of crystalline structure on the EXAFS pattern. Analysis of results of magnetic measurements 
(dc magnetization and ac susceptibility) showed a combination of superparamagnetic blocking 
and collective superspin glass freezing.  The interpretations of the results indicate that 
the presence of Pt shell is the key factor in observed magnetic behavior. Specifically, the 
Pt shell, in combination with the small size of Co-based core, can polarize (or frustrate) 
the spins at the surface and in the near-surface regions of antiferromagnetic CoO nanoparticles. 
Such cooperation of two different processes with completely different spins dynamics in Co-based 
nanoparticles has not been previously reported and is an important finding of this work.  
Usually the existence of shell layers with different 
spin dynamics leads to an exchange bias effect. However, in the sample under study, 
no bias mechanism was observed, thereby indicating the mechanism of spin glass freezing 
that leads to the SSG state is mainly influenced by the strong dipolar magnetic interactions of superspins.

The presentation of our work starts by providing the details of the synthesis protocols used to 
fabricate the samples, and the subsequent characterization methods used to explore the chemical 
composistion and morphology.  Next, the magnetic investigations and results are described, while 
an extentend discussion of the anlaysis of all of the data are given.  Finally, the paper 
concludes with a summary of our findings.   

\section{EXPERIMENTAL DETAILS}
\subsection{Sample Preparation}

The CoO-Pt core-shell nanoparticles were synthesized using the reverse micelle 
concept,\cite{800568,Lin200126,Lee2003} based on dissolution of a cationic surfactant in an organic 
solvent and the formation of spherical reverse micelle aggregates. All chemicals were purchased from 
Aldrich and Sigma and used without further purification. The reverse micelle solutions were prepared 
using cetyltrimethylammonium bromide (CTAB) as the surfactant with octane as the oil phase, while   
1-butanol was used as a co-surfactant helping to stabilize the micelle solutions. Water solutions of  
CoCl$_{2}$, NaBH$_{4}$ and H$_{2}$PtCl$_{6}$ were used for the formation of initial droplets in the 
reverse micelles, whose reaction lead to the final nanoparticles. The size of the particles 
(water droplet size) was controlled by adjusting of the water-to-surfactant molar ratio  
[H$_{2}$O/CTAB] = 5. The Co based particles formed the cores coated by a  
non-magnetic Pt surface layer.  This process prevents the agglomeration of particles and   
is effective in preparing particles with metal cores and different shells.

\subsection{Characterization}
High-resolution transmisson electron microscopy (HRTEM) and electron 
diffraction analysis were performed with a JEOL JEM 3010 
transmission electron microscope operated at 300~kV (LaB$_{6}$ cathode). Copper grids coated with a 
holey carbon support film were used to prepare samples for the TEM studies. Powdered samples 
were dispersed in ethanol, and the suspension was treated in an ultrasonic bath for 10~min.

The structure of the nanoparticles was investigated by means of the angular dispersive 
x-ray diffraction (XRD) and x-ray absorption spectroscopy (XAS).  The XRD measurements using 
high-energy photons were performed at the wiggler beamline BW5 at HASYLAB/DESY (Hamburg, Germany). 
The wavelength of the radiation was set to 0.12398~\AA{}, which corresponds to the beam energy of 
100~keV. Thin-walled (20~$\mu$m) quartz capillaries having diameters of 2~mm were used for the XRD 
measurements. A powder sample was illuminated for 50~s with the well collimated beam having the 
cross-section of $1 \times 1$ mm$^{2}$. A LaB$_{6}$ standard was used to calibrate the sample-to-detector 
distance along with the tilt of the imaging plate relative to the beam path and 
to determine the instrumental broadening. Diffraction patterns were collected in 
transmission mode using a MAR345 image plate detector. 
Two dimensional XRD patterns were radially integrated using the FIT2D program.\cite{doi:10.1080/08957959608201408}

The local atomic arrangement was investigated by x-ray absorption spectroscopy (XAS), which provides 
complementary information to XRD measurements though offering better chemical sensitivity. The XAS 
measurements were conducted at the bending magnet experimental stations C and X1 at HASYLAB/DESY. 
The fine oscillations of the linear absorption coefficient $\mu$(E) were measured at the Co~$K$ edge 
(7709~eV) and at the Pt~$L_{3}$ edge (11564~eV) in transmission mode using a fixed exit 
double-crystal Si (111). In order to optimize absorption signal, powder nanoparticles (7.1 mg) 
were uniformly dispersed with cellulose powder (200 mg). Compressing the mixture using a hydraulic 
press yielded a sample pellet with diameter and height of 13 and 3 mm, respectively. The pellet was 
placed behind the first ionization chamber and was illuminated by the incoming beam having the 
cross-section $5 \times 1$ mm$^{2}$. The energy calibration was performed simultaneously with the 
sample measurement by putting the corresponding reference Co or Pt foil behind the second ionization chamber. 
Experimentally measured x-ray absorption cross section $\mu$(E) was analyzed by the standard procedures 
of data reduction described elsewhere\cite{0022-3727-34-2-309,0953-8984-19-17-176215} using the 
program VIPER.\cite{0022-3727-34-2-309,Viper} Firstly, the EXAFS signal $\tilde{\chi}(k)$ was extracted, 
weighted by $k^{2}$, and subsequently Fourier transformed (FT) into the real space of interatomic 
distances. Then the main peak of the FT-$|k^{2}\,\tilde{\chi}(k)|$ signal was separated by applying the so-called 
Hanning window function with the coefficient $A = 0.01$. The resulting data were then 
inverse Fourier transformed back into $k$-space, and these filtered data only contain information 
about the atoms nearest to the absorbing ones (first coordination shell).  Finally, fitting by an 
appropriate model yields structural information about the coordination number $N$, the interatomic 
distance $r$, and the Debye-Waller factor $\sigma$. The backscattering amplitude 
$F_{i}(k)$ and phase shift $\tilde{\chi_{i}}(k)$ functions, necessary for computation of the EXAFS 
signal from the model, were calculated using the FEFF 6 code.\cite{PhysRevB.58.7565}

\begin{figure}[t]
\includegraphics[trim = 0 -5mm 0 0, clip, width=8.6cm]{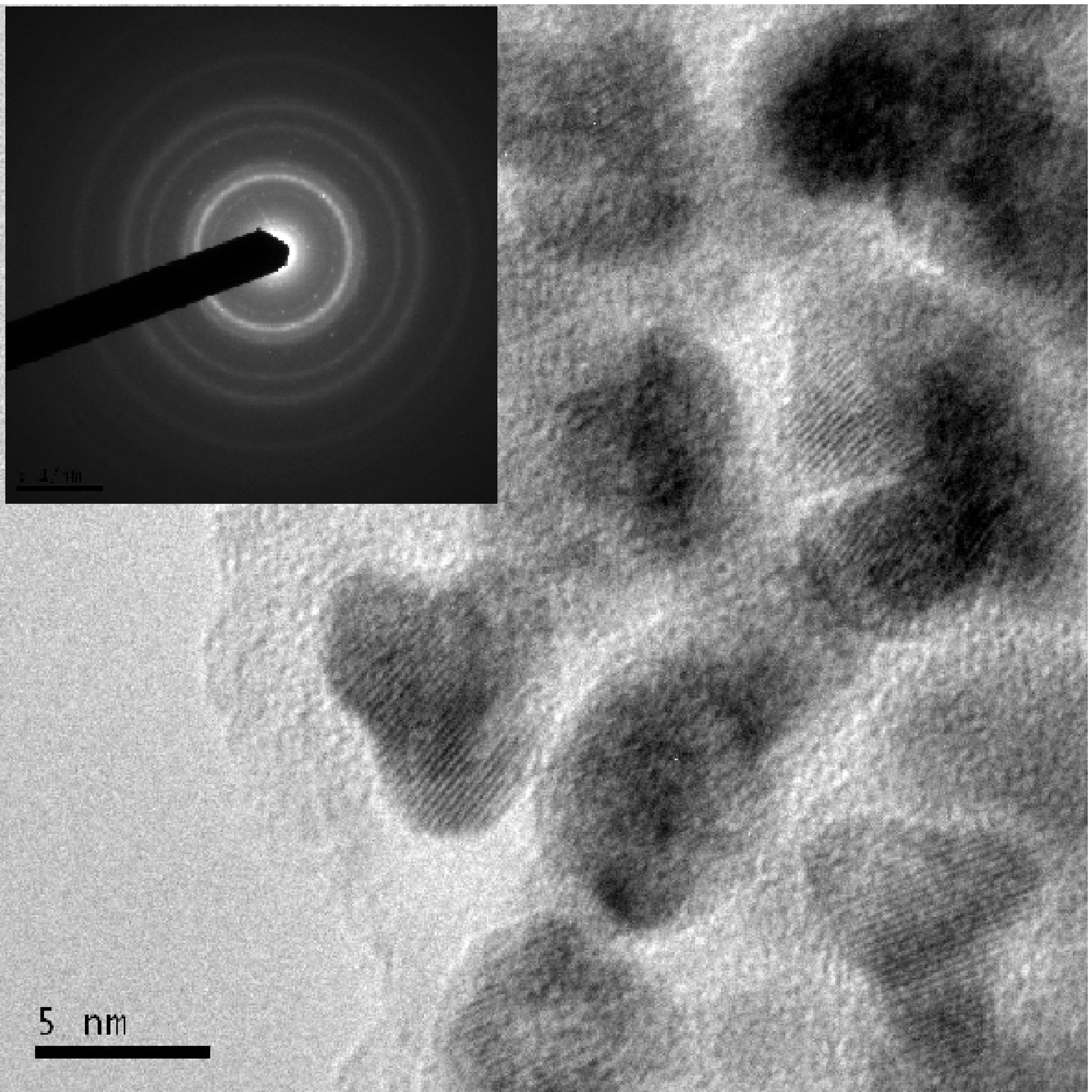}
\includegraphics[width=8.6cm]{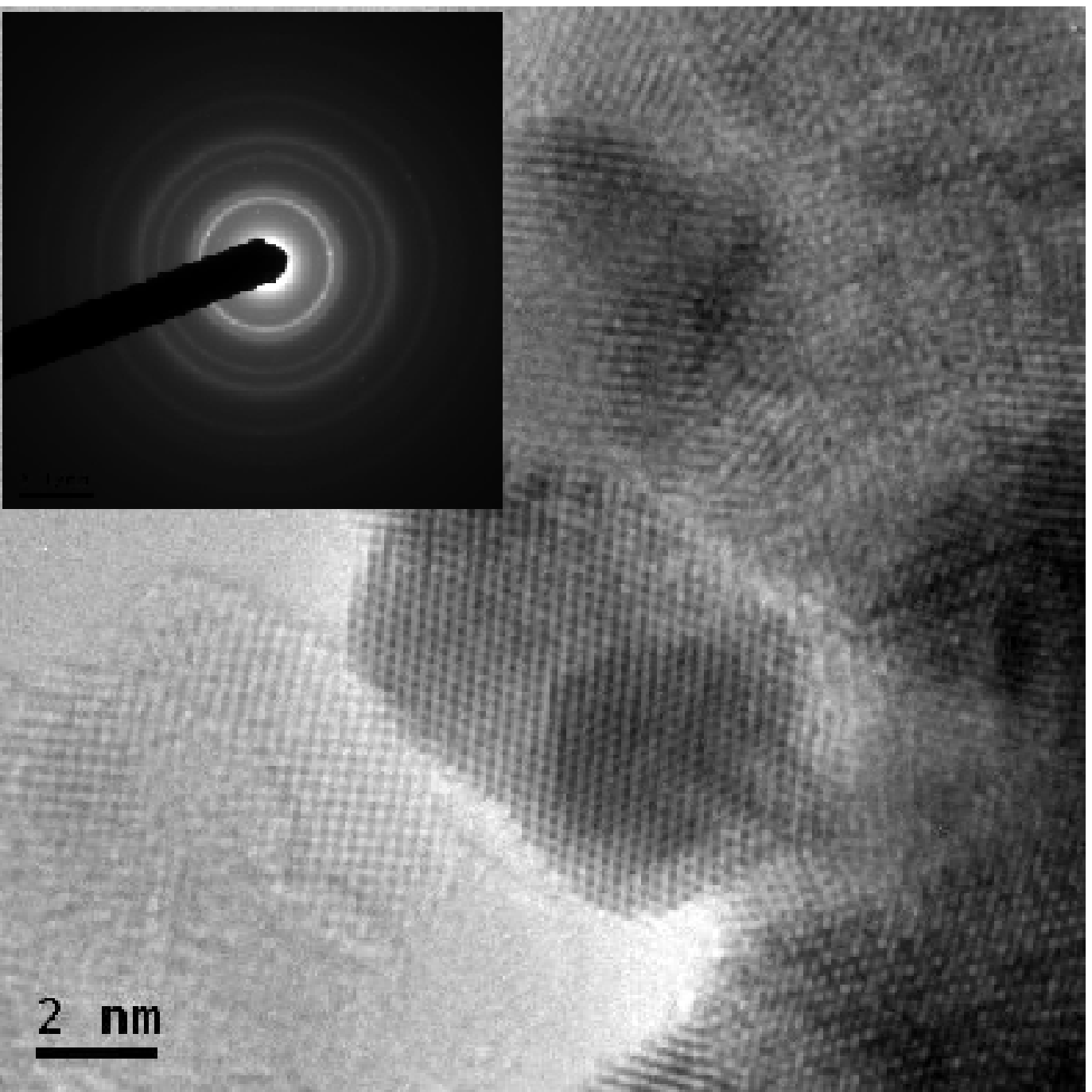}
\caption{\label{fig1}HRTEM micrographs and diffraction patterns of from two different 
samples of the CoO-Pt core-shell nanoparticles.}
\end{figure}

Magnetic measurements were performed on a commercial superconducting quantum interference device (SQUID) 
magnetometer (Quantum Design MPMS XL5) over a range of temperatures (2~K $\leq T \leq 300$~K) and in 
applied static magnetic fields up to 50~kOe. A sample with mass 6.9~mg was placed in a plastic capsule 
that was supported by a plastic sample holder. The diamagnetic contribution of the capsule and holder 
are insignificant compared with the large magnetic signal of the sample, so no correction was necessary. 
For $T \leq 150$~K, the complex ac magnetic susceptibility, $\chi'(T,\nu) + i\chi''(T,\nu)$, was 
recoreded by the same instrument using an ac magnetic field of 2.5~Oe in the frequency interval 
1~Hz $\leq \nu \leq 1$~kHz while no dc magnetic field was applied.
	
\section{RESULTS AND DISCUSSION}
\subsection{Structural properties}

Typical micrographs, showing the size and shape of the resulting CoO-Pt 
core-shell nanoparticles, are shown in Fig.~\ref{fig1}, and particle size (log-normal) distribution 
analysis yielded diameters of $(4.0 \pm 0.2)$~nm.  Although HRTEM evidence of the core-shell morphology 
was not resolved, this result is consistent with the findings of 
Park and Cheon,\cite{doi:10.1021/ja0156340} whose HRTEM images showed only smooth and homogenous 
boundaries for solid-solutions and core-shell CoPt nanoparticle samples. 
However, the core-shell nature of our CoO-Pt nanoparticles was resolved by our EXAFS experiments, 
\emph{vide infra}. Electron diffraction patterns, shown as insets in Fig.~\ref{fig1}, clearly indicate 
the crystalline character of the sample and long-range structural order.

\begin{figure}[t]
\begin{center}
\includegraphics[trim = 10mm 12mm 45mm 20mm, clip,width=8.6cm]{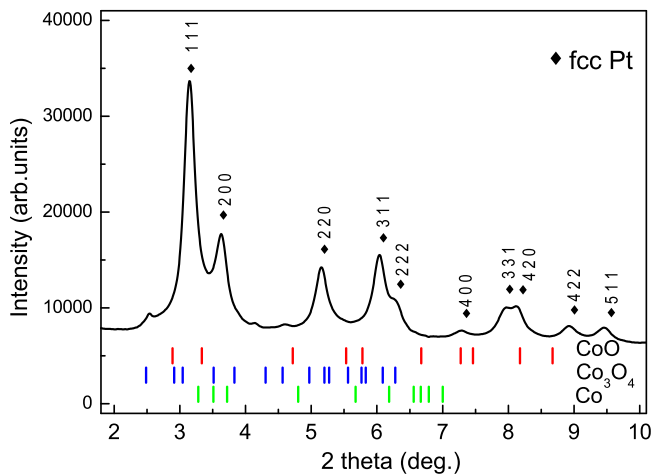}
\end{center}
\caption{\label{fig2} (Color online) XRD pattern, of CoO-Pt core-shell nanoparticles, obtained 
using synchrotron radiation with a wavelength 0.12398~\AA.  Peak positions associated with 
fcc Pt, CoO, Co$_3$O$_4$, and Co are designated, and a full discussion of the results 
is given in the text.}
\end{figure}

The XRD pattern of our CoO-Pt core-shell nanoparticles is shown in Fig.~\ref{fig2}, 
where the relatively broad Bragg peaks indicate the nanocrystalline nature of the sample. 
While the analysis suggests the presence of the fcc Pt phase (PDF 40-802), the CoO phase 
can not be resolved unambiguously due to overlapping of the anticipated CoO peaks with 
the ones associated with the Pt phase. These observations are typical 
for core-shell nanostructures of iron or cobalt fine nanoparticles coated with Au 
or Pt.\cite{Carpenter2001,JACS2004} So, the CoO phase in our particles was resolved by our 
EXAFS and XANES studies, \emph{vide infra}.  Nevertheless, the XRD can be used to extract the 
average grain size from the line broadening by applying the Scherrer formula.\cite{Scherrer} 
More specifically, Bragg peaks at (111), (200), and (220) were fitted to Gaussian functions 
to extract values for the FWHM, which were corrected for the instrumental broadening. 
Finally, the anlaysis yielded values for the average size of nanoparticles, $(3.4 \pm 0.4)$~nm, and  
the lattice parameter, $(3.908 \pm 0.004)$~\AA{}.

In order to provide experimental evidence for the CoO-Pt core-shell morphology of our samples, 
XAS measurements at the Pt $L_3$ and Co $K$ edges were performed. A great advantage of XAS 
is its chemical sensitivity that provides a mircoscopic probe of the atomic neighborhood of 
a selected atomic constituent. Generally, a signal obtained from XAS consists of two parts: 
(\emph{i}) the part near the absorption edge, literally named XANES 
(X-ray absorption near-edge structure) and (\emph{ii}) EXAFS 
(extended X-ray absorption fine structure) that starts $30-60$~eV above an absorption edge 
and persists up to about 1~keV beyond the edge by revealing specific oscillations of the 
absorption coefficient. In ohter words, XANES can provide information 
about the electronic configuration and is sensitive to the oxidation state of the absorbing atom, while 
EXAFS contains information about coordination environment of the absorbing atom.

\begin{figure}[ht]
\includegraphics[trim = 0 -5mm 0 0, clip, width=8.6cm]{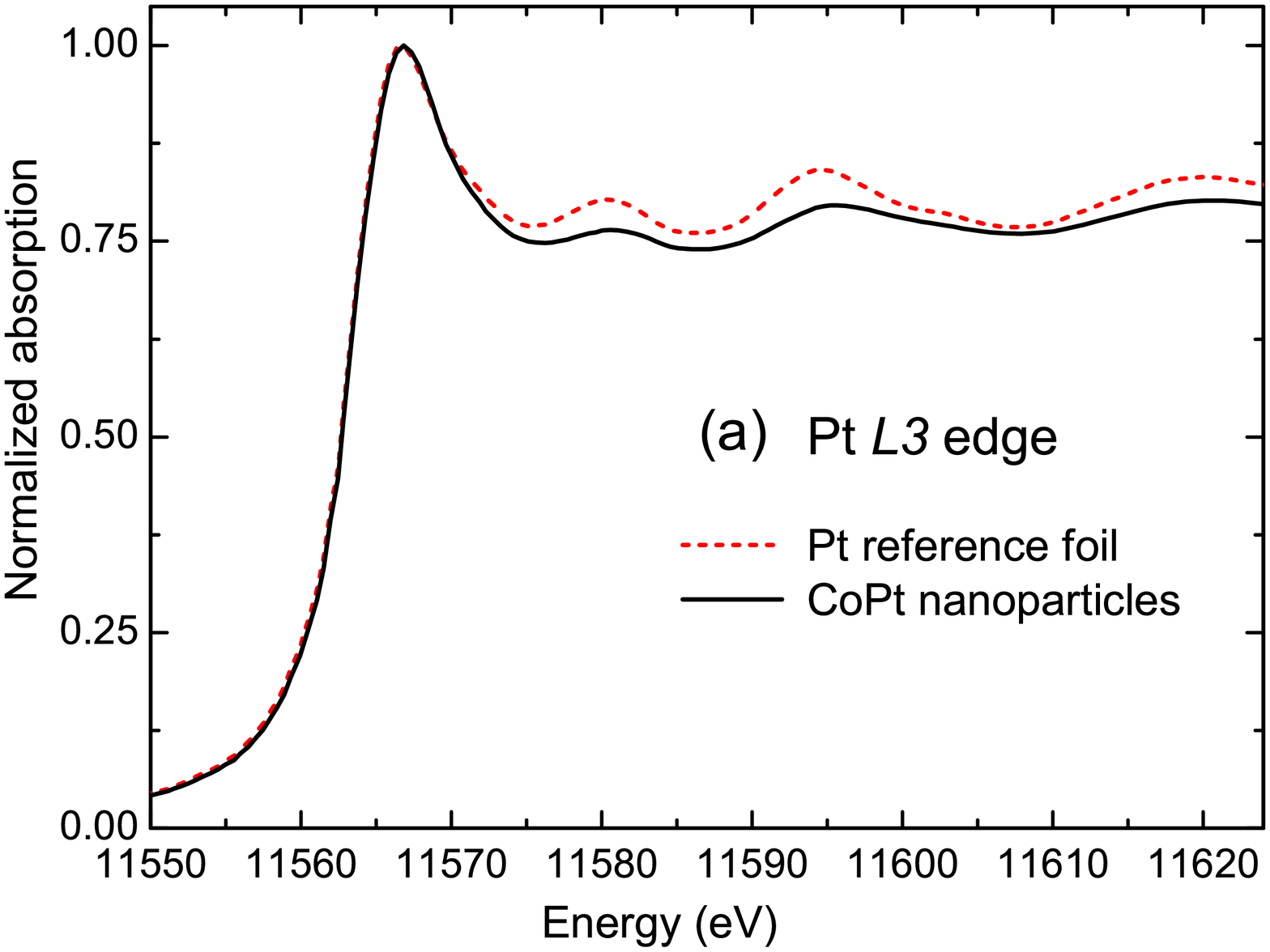}
\includegraphics[width=8.6cm]{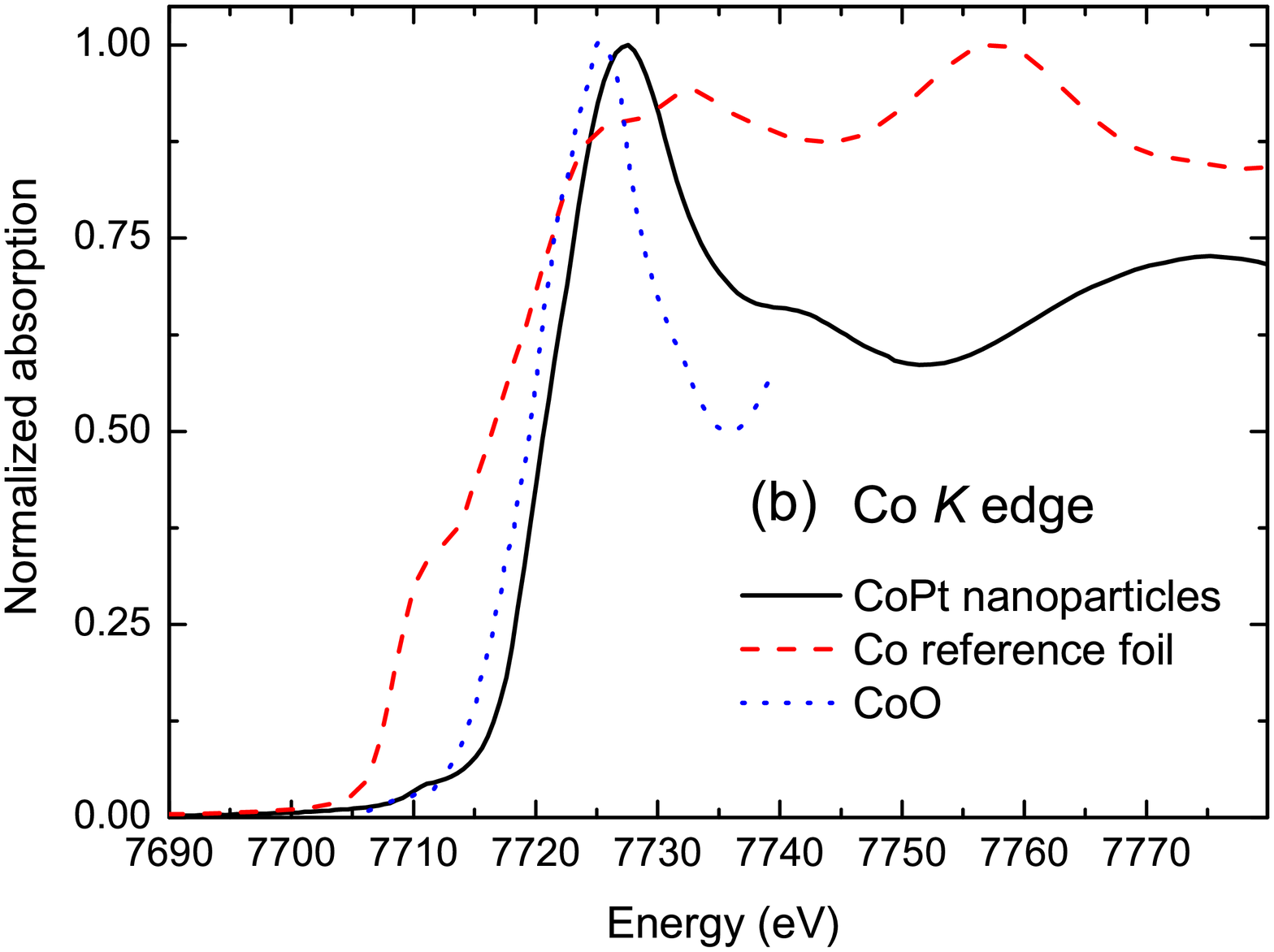}
\caption{\label{fig3}(Color online) Normalized (a) Pt $L_{3}$ and (b) Co~$K$ absorption edges of CoO-Pt 
core-shell nanoparticles (solid line) and Pt and Co reference foil (dash line), respectively. 
Dotted line represents the CoO reference.\cite{Cheng2004122}}
\end{figure}

The XANES spectra for nanoparticles obtained at the Pt $L_{3}$ and Co $K$ edges are shown 
Fig.~\ref{fig3}, together 
with XANES spectra obtained from  pure Pt and Co reference foils (with micrometer size), respectively. 
In the case of the Pt $L_{3}$ edge, the XANES signal coming from CoO-Pt core-shell nanoparticles is 
practically identical to the signal from Pt reference foil. This behavior indicates the Pt atoms 
in the nanoparticles have comparable electronic configurations and the same oxidation states (exactly 
the same position and the shape of the absorption edge) as Pt atoms in the reference foil. 
The situation at Co $K$ edge is completely different, and the XANES signal of CoO-Pt core-shell particles has no common features with XANES extracted from the Co reference foil, as documented in Fig.~3~(b). 
It implies that Co atoms of studied nanoparticles feel diverse electronic surrounding as Co 
atoms in the reference foil. Shifting and mainly shape changing of the absorption edge suggests 
different oxidation state of cobalt atoms, specifically a CoO phase. 
These results observed at the Co absorption edge are in accordance with the observations of 
Cheng and co-authors,\cite{Cheng2004122} who used XANES and EXAFS methods for a detailed 
structural study of Co nanoparticles with different sizes (3~nm, 5~nm, and 12~nm) prepared 
under anaerobic conditions. Their XANES results showed that the pattern from the 12~nm Co particles 
closely resembled the data from the Co foil at the 7712~eV edge,  while the pattern for the 
5~nm particles had a much smaller shoulder and the 3~nm particles showed almost no signature. 
In fact, the data\cite{Cheng2004122} from the 3~nm particles are closest to signatures expected 
from a CoO standard. In other words, the smallest Co particles (3~nm) possess a 
surface layer of oxygen eventhough the synthesis was realized in anaerobic conditions.\cite{Cheng2004122} 
The results from our study are consistent with this trend, where the presence of CoO was identified 
in the XANES data. 

\begin{figure}[t]
\includegraphics[trim = 0 -5mm 0 0, clip, width=8.6cm]{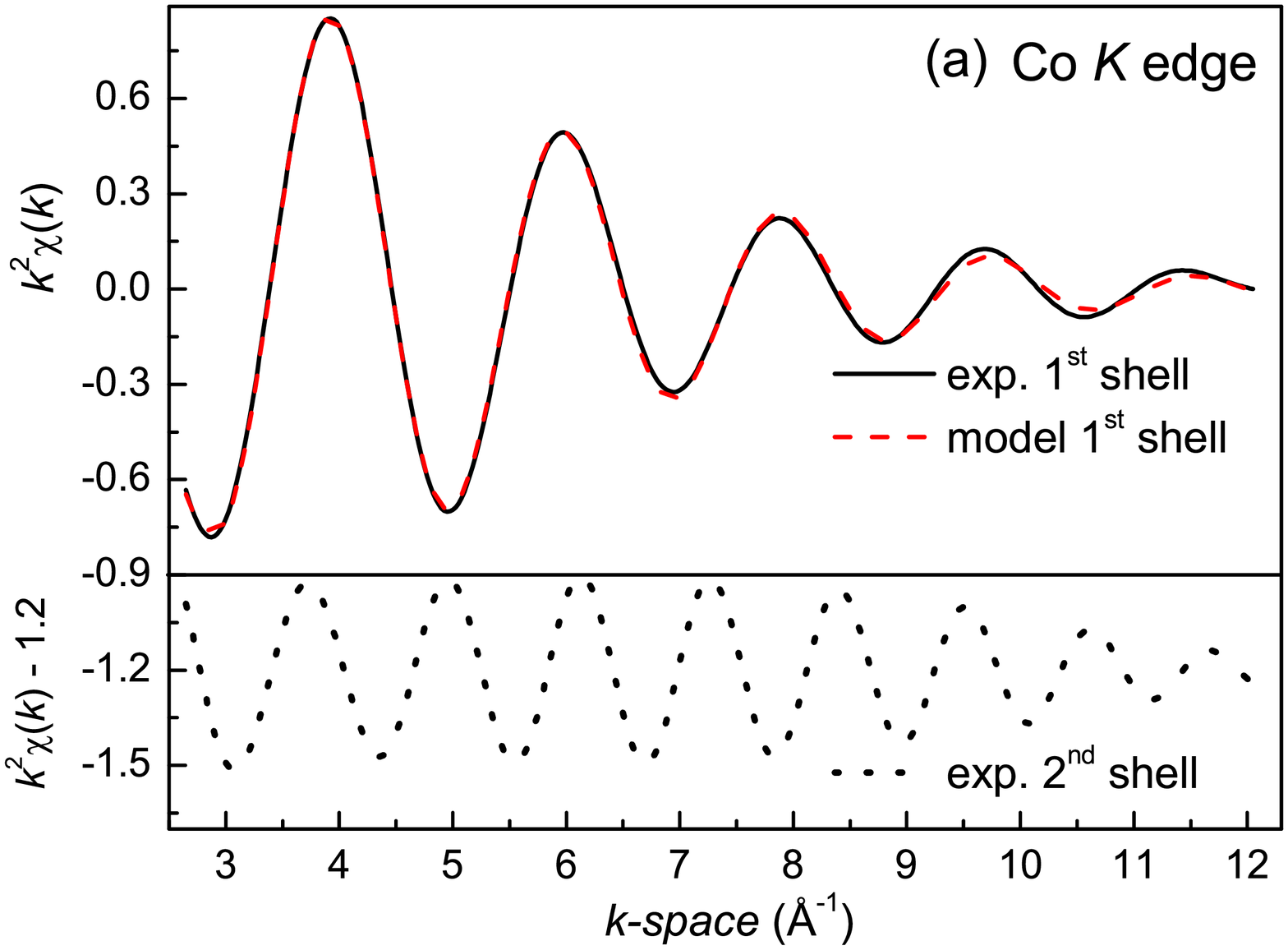}
\includegraphics[width=8.6cm]{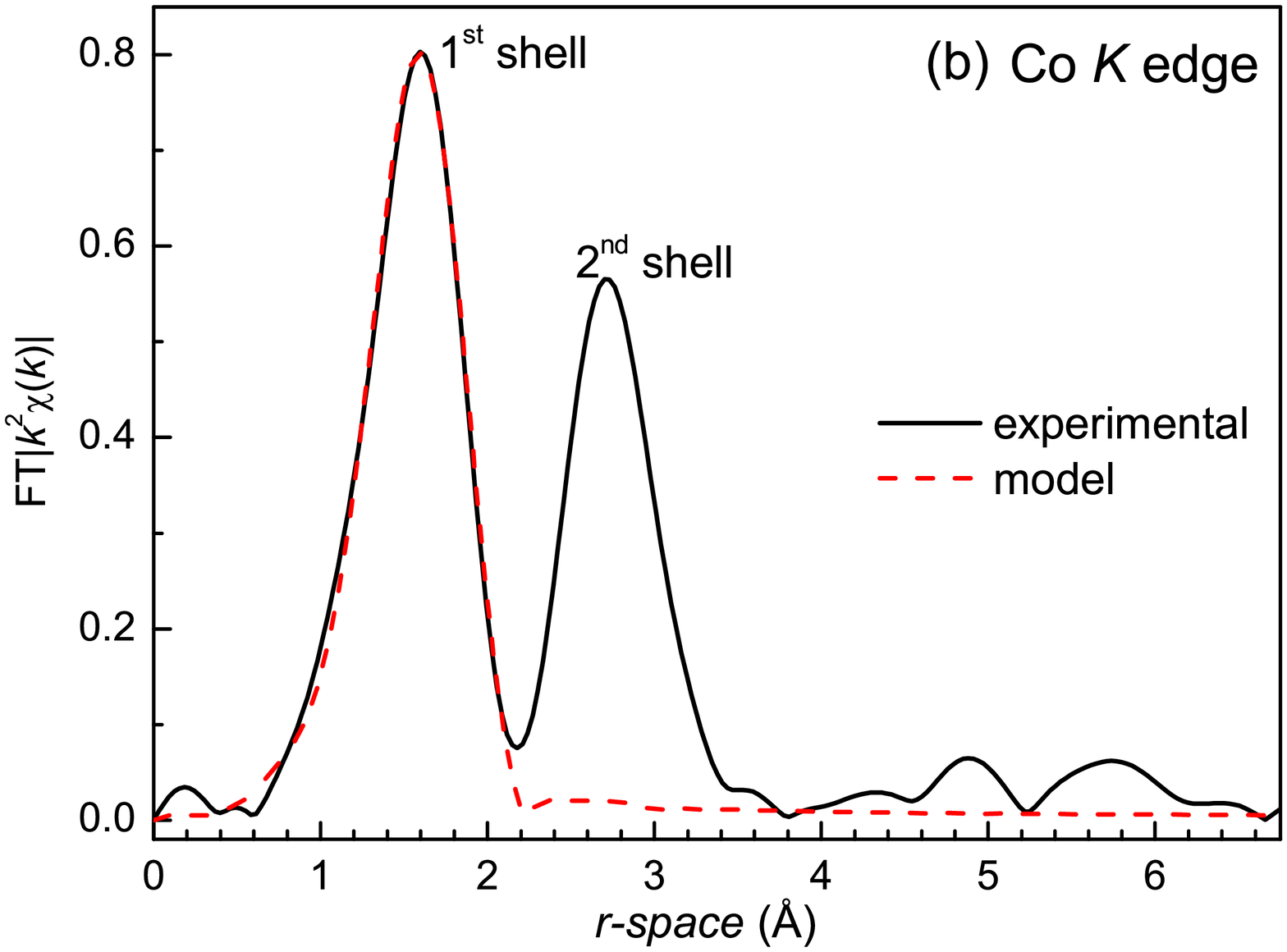}
\caption{\label{fig4}(Color online) Experimental EXAFS data (solid line) at Co $K$ 
absorption edge and best fit result (dash line) for CoO-Pt core-shell nanoparticles. The comparison shows (a) 
back Fourier-transform in $q$-space and (b) magnitude of Fourier-transform in $r$-space.}
\end{figure}

\begin{figure}[t]
\includegraphics[trim = 0 -5mm 0 0, clip, width=8.6cm]{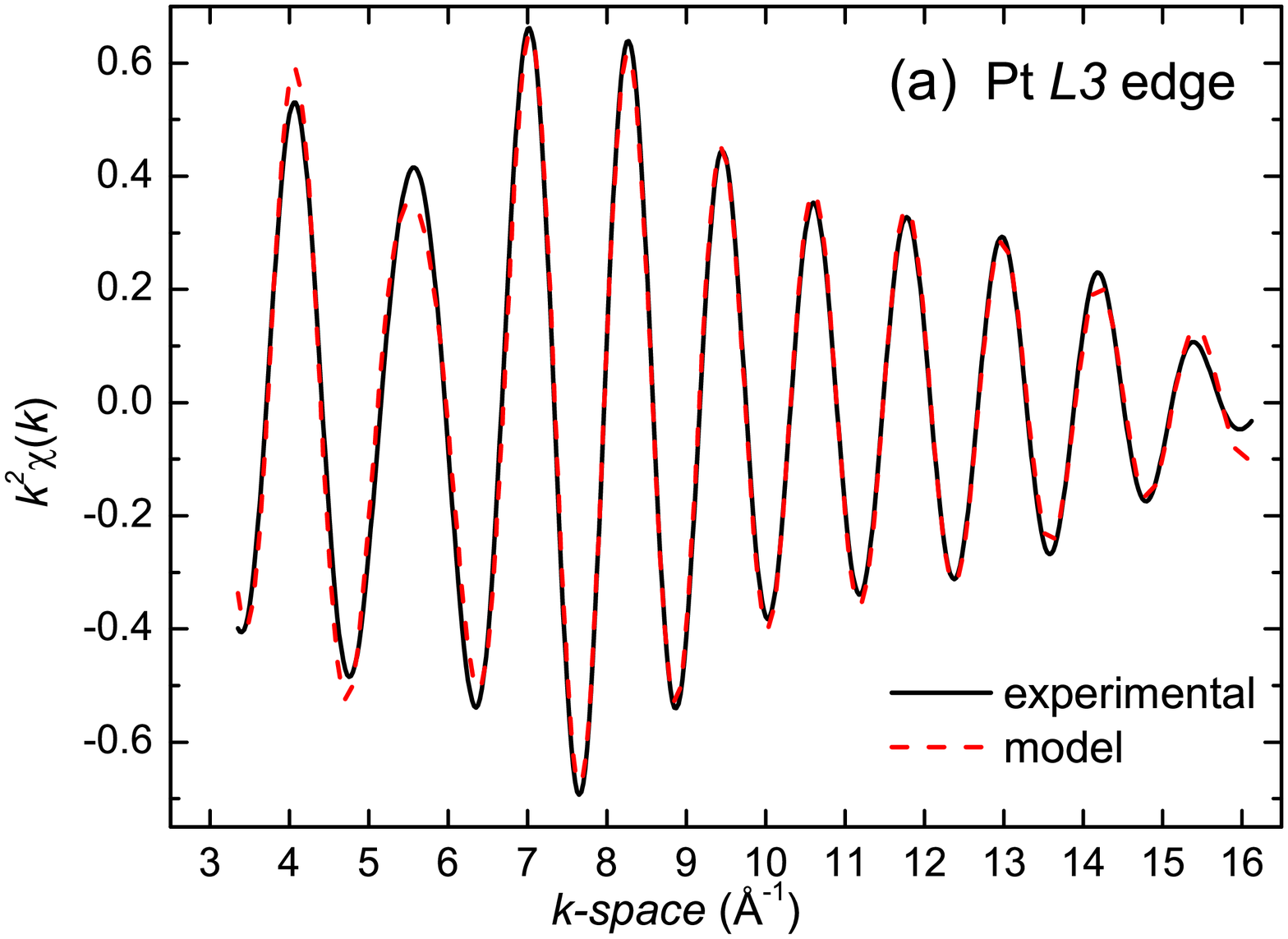}
\includegraphics[width=8.6cm]{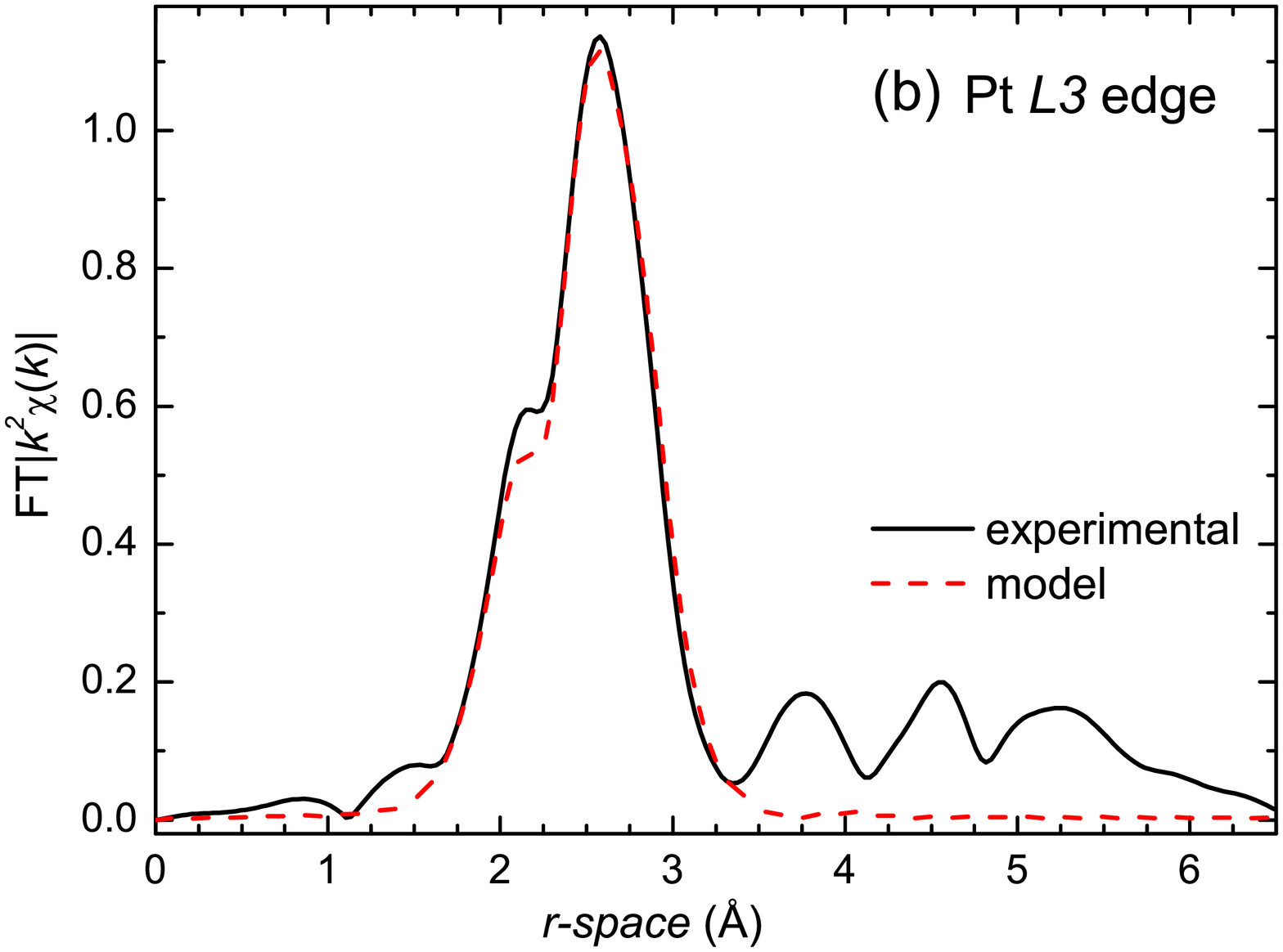}
\caption{\label{fig5}(Color online) Experimental EXAFS data (solid line) at the Pt $L_{3}$ absorption edge and best 
fit result (dash line) for CoO-Pt core-shell nanoparticles. The comparison shows 
(a) the inverse Fourier-transform in $q$-space and 
(b) the magnitude of Fourier-transform in $r$-space.}
\end{figure}

Additional insight about the structure of our samples was obtained from the EXAFS data, 
shown in Figs.~\ref{fig4} and \ref{fig5}.  Immediately obvious is that the Fourier transformation 
of the weighted signal at the Co $K$ edge 
is characterized by two well separated peaks situated at 1.6~\AA{} and 2.7~\AA{}, 
see Fig.~\ref{fig4}(b). These peak positions corresponds roughly to the most probable 
interatomic distances between an absorbing atom and its nearest neighboring atoms. 
It is important to note that the interatomic distances extracted in this manner are usually 
underestimated, so one has to apply phase shift corrections to obtain reasonable values. 
Nevertheless, the location of the first peak at relatively low $r$ values suggests the 
presence of atoms, with a small atomic radius, surrounding the Co. Consequently, 
an inverse Fourier transformation of the first peak can be performed, as described in the 
preceeding section, with  0.61~\AA{} $\leq r \leq 2.22$~\AA{}, and fitted by a 
shell of oxygen atoms.  The results of the fit are compared to the data in 
Fig.~\ref{fig4}(a), and the values of the fitting parameters are 
$N_{\mathrm{Co-O}} = 6.6 \pm 1$, $r_{\mathrm{Co-O}} = 2.05 \pm 0.03$~\AA{}, 
$\sigma^{2} = 0.0105 \pm 0.0005$~\AA{}$^{2}$ with $R = 0.0460$. 
The same procedure was applied to the second peak, which contains 
information about the second coordination shell, where the analysis is 
restricted to 2.19~\AA{} $\leq r \leq 3.47$~\AA{}. 

To unambiguously confirm that the external shell of nanoparticles consists exclusively of 
Pt atoms and does not contain any Co atoms, the Pt $L_3$ EXAFS signal, Fig.~\ref{fig5} was analyzed using two 
different models. The initial fitting was performed for a cluster of 12 Pt atoms (the 
first coordination shell) surrounding an absorbing Pt atom with positions corresponding 
to the atomic positions in fcc Pt phase, as motivated by the XRD results. 
The second model is a modification of the first one, but with 4 of the 12 Pt atoms were 
replaced by Co atoms. This model represents the Pm3m CoPt$_{3}$ phase (PDF 299-499). 
We have found that including Co atoms in the first coordination shell strongly decreases 
the fit quality if the coordination number $N_{\mathrm{Pt-Co}}$ is constrained to the values 
between 2 and 4. When leaving $N_{\mathrm{Pt-Co}}$ unconstrained, its value tended to zero, 
meaning that there is essentially no contribution from Co atoms in the first coordination 
shell.  Therefore, we conclude the shell surrounding the Co core consists exclusively of 
Pt atoms. As can be seen from Fig.~\ref{fig5}, the final fit of the experimental data shows 
nice agreement with the first model based on a pure fcc Pt phase, where the extracted 
structural parameters  
$N_{\mathrm{Pt-Pt}} = 10.3 \pm 1$, $r_{\mathrm{Pt-Pt}} = 2.73 \pm 0.01$~\AA{}, 
$\sigma^{2} = 0.008 \pm 0.0005$ \AA{}$^{2}$ with $R = 0.0916$.  
To summarize the structural studies based on results of XRD, XANES 
and EXAFS experiments, our samples consist of core CoO particles with pure Pt shells.

\begin{figure}[t]
\includegraphics[width=8.6cm]{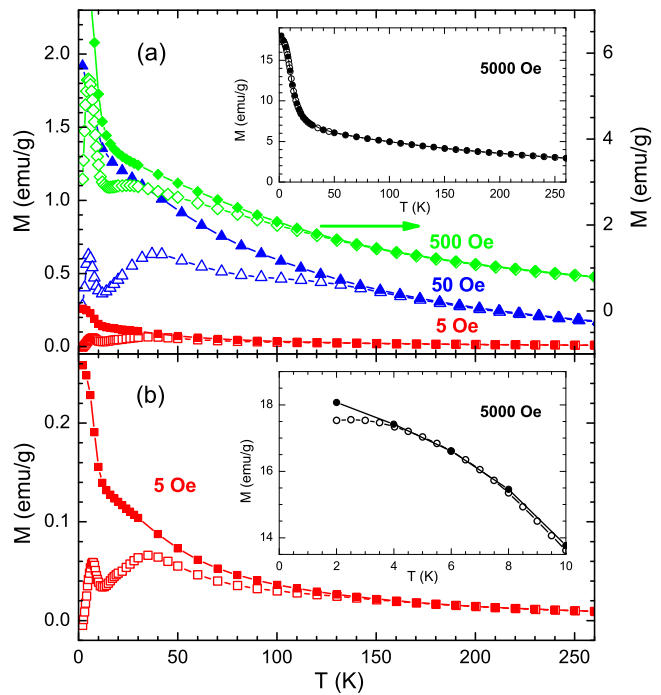}
\caption{\label{fig6}(Color online) Temperature dependence of the dc magnetization measured in fields up to 5000 Oe 
after ZFC (open symbols) and FC (solid symbols). The full temperature range is shown in (a), while 
(b) provides an expanded views of the 5 Oe and 5000 Oe data.  Table~\ref{table1} provides a tabulation of the features 
discussed in the text.}
\end{figure}

\subsection{dc magnetization}
\subsubsection{Temperature dependence}

The temperature dependences of the zero-field-cooled (ZFC) and field-cooled (FC) dc-magnetizations, 
$M_{\mathrm{ZFC}}$ and $M_{\mathrm{FC}}$,  in low static 
fields ($\leq 5000$~Oe) are shown in Fig.~\ref{fig6}.  In these low fields, the magnetic response 
bifurcates at a temperature, $T_{\mathrm{irr}}$, exhibiting irreversibility at lower 
temperatures.  In addition, a striking feature is the presence of two discrete ZFC maxima in 5 Oe, 
suggesting two blocking or freezing temperatures, $T_{\mathrm{M}1}$ and $T_{\mathrm{M}2}$. 
The field dependences of these three characteristic temperatures are tabulated in 
Table~\ref{table1}.

\begin{table}[b]
\begin{center}
\begin{tabular}{|r|c|c|c|}
\hline
$\;B$ (Oe) $\;$ & $\;\;T_{\mathrm{M}1}$ (K) $\;\;$ & $\;\;T_{\mathrm{M}2}$ (K) $\;\;$ & $\;\;T_{\mathrm{irr}}$ (K) $\;\;$ \\
\hline
5 $\;\;$ & 6 & 36 & 168 \\
\hline
50 $\;\;$ & 5 & 37 & 142 \\
\hline
500 $\;\;$ & 4 & 29 & 115 \\
\hline
5000 $\;\;$ & 3 & $-$ & $\;\;$ 3 \\
\hline
\end{tabular}
\caption{\label{table1}Field dependences of the temperatures of the two maxima in the 
ZFC data and the irreversity temperatures of the data shown in Fig.~\ref{fig6}.}
\end{center}
\vspace{-0.0cm}
\end{table}

\begin{figure}[ht]
\includegraphics[width=8.6cm]{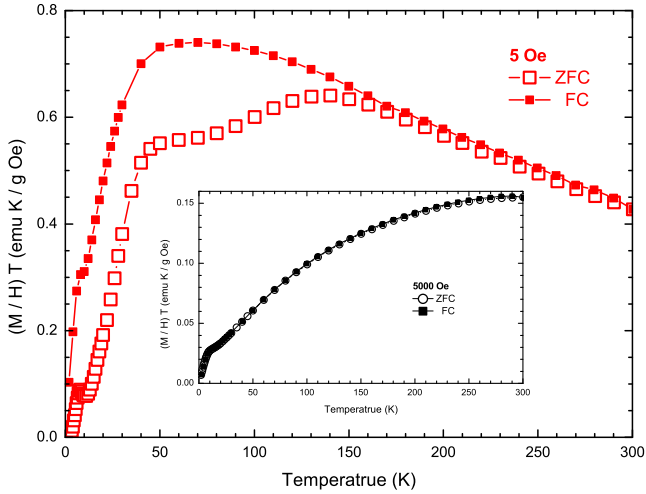}
\caption{\label{fig6bis}(Color online) Temperature dependence of the dc magnetization 
of the CoO-Pt core-shell nanoparticles measured in field 5 Oe when using ZFC 
and FC protocols.  The inset shows $M_{\mathrm{ZFC}}$ and $M_{\mathrm{FC}}$ 
curves measured in 5000 Oe. These data sets in 5 Oe and 5000 Oe show the influence 
of the magnetic field on the shape of curves.}
\end{figure}

Another significant feature of the data shown in Fig.~\ref{fig6} is the behavior of $M_{\mathrm{FC}}$, 
which shows a progressive increase with decreasing temperature below  $T_{\mathrm{M2}}$ and a sharp rise 
below $T_{\mathrm{M1}}$.  The sharp increase at the lowest temperatures suggests an additional magnetic 
contribution and may be related to the collective freezing of the disordered 
spins in the interface ``spin'' layer between antiferromagnetic CoO core and non-magnetic Pt shell.

The presence of two maxima in $M_{\mathrm{ZFC}}(T)$ data acquired in low fields was also reported by 
Winkler \emph{et al.}\cite{PhysRevB.72.132409,0957-4484-19-18-185702} and Thota and Kumar\cite{Thota20071951} 
for NiO core-shell nanoparticles with antiferromagnetic cores. They associated this peculiar behavior 
with the freezing of the magnetically ordered region in the surface shell. In addition, 
Zhang \emph{et al.}\cite{Zhang2003111} interpreted the anomalous properties of antiferromagnetic 
CoO nanoparticles in terms of a core-shell model, 
where the ferromagnetic portion is attributed to the increase of the uncompensated moments at the surface 
resulting from the reduced coordination of surface spins.  Furthermore, the results of 
Zysler \emph{et al.}\cite{Zysler2003233} and Biasi \emph{et al.}\cite{PhysRevB.71.104408} show on the 
important role of the surface anisotropy in determination of the anomalies of dc magnetization of 3 nm 
sized magnetic nanoparticles. Likewise, Dutta \emph{et al.}\cite{Dutta2009} employed a core-shell model, which assumes the coexistence of an ordered core with disordered surface spins, to explain unconventional magnetic properties of nominally 4~nm iron oxide nanoparticles. 

\begin{figure}[ht]
\includegraphics[width=8.6cm]{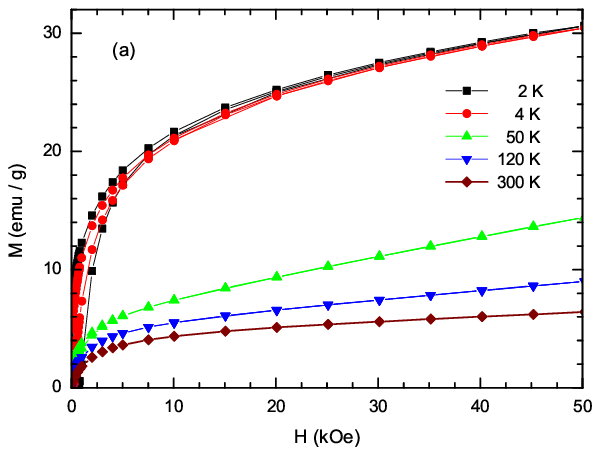}
\includegraphics[width=8.6cm]{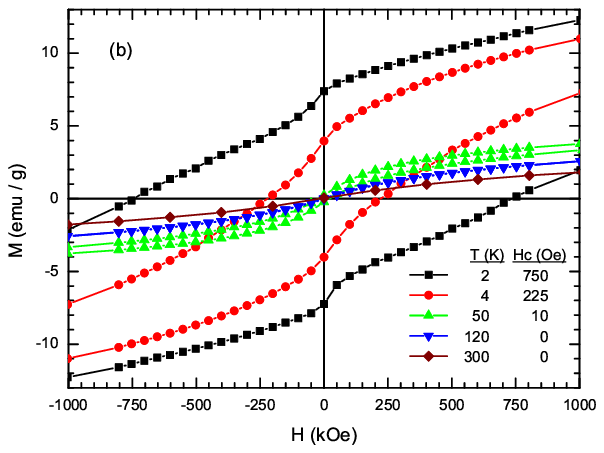}
\caption{\label{fig7}(Color online) Field dependences of the isothermal magnetization measured at 
several different temperatures.  The first $M$ vs.~$H$ quadrant is shown in (a).  An expanded view 
of the low field region of the hysteresis loops is provided in (b), and the values of the coercive 
fields are given in the legend.}
\end{figure}

The existence of these two, low-temperature maxima in low magnetic fields begs the question as to the 
nature of the magnetic state at high temperatures.  In an attempt to resolve this issue, magnetic 
susceptibility $(\chi = M/H)$ times temperature versus temperature plots were constructed, and the results 
for the data sets in 5~Oe and 5000~Oe are shown in Fig.~\ref{fig6bis}.  The data for these two fields 
show contrasting behavior, as the data in 5~Oe indicate a ferromagnetic trend for 
150~K $\lesssim T \leq 300$~K while the results in 5000~Oe indicate an antiferromagnetic trend for 
nearly all temperatures. In other words, the signatures present in the static magnetic responses appear to 
possess a complex interplay of temperature and magnetic field, so the results of isothermal $M(H)$ studies 
will be presented before a full discussion is presented at the end of this subsection.

\subsubsection{Field dependence}
After ZFC, the isothermal $M(H)$ loops were obtained by sweepting to $\pm \, 50$~kOe at several temperatures, 
Fig.~\ref{fig7}.  For $T \geq 120$~K, the magnetization curves are reversible and no 
coercivity was detected, whereas for $T \leq 50$~K, hysteresis 
was observed, and the values of the coercive fields are listed in the legend of Fig.~\ref{fig7}(b).  
For $T \geq 50$~K, it is noteworthy that a temperature-dependent paramagnetic constribution is 
observed as a straight-line for 30~kOe $\leq H \leq 50$~kOe.  This observation suggests
a modified Langevin formalism, which has been applied to other systems for $T_{\mathrm{B}} \lesssim T \lesssim T_{\mathrm{N}}$,\cite{PhysRevB.55.R14717,PhysRevB.61.3513,PhysRevB.64.132410,PhysRevB.69.054425} might be applied to the $M(H)$ data sets at 50~K and 120~K, because neither a classical Langevin function\cite{Kim200130} nor a weighted sum of Langevin functions\cite{Tadic200912,PhysRevB.66.104406} plausibly  simulated the $M(H)$ data. The modified Langevin analysis provides coarse estimates of the nanoparticle magnetic moments of 1500~$\mu_{\mathrm{B}}$ at 50~K and 2000~$\mu_{\mathrm{B}}$ at 120~K, where, presumably, the difference between these two values arises from thermal variations of the antiferromagnetic interactions.  

Using the data reported by Silva \emph{et al.},\cite{PhysRevB.82.094433} the moments at 
50~K and 120~K might be expected from 4~nm diameter 
nanoparticles of antiferromagnetically ordered CoO.  However, the N\'{e}el temperature of 
nanoparticles of this size are expected to be significantly less than 240~K.\cite{PhysRevB.82.094433} 
So, the temperature of the maximum value of $\partial(M/H)T/\partial T$, for the ZFC data set with subsequent measuring in 5~Oe, Fig.~\ref{fig7}, is near 145~K and provides an estimate 
of $T_{\mathrm{N}}$.\cite{PhysRevB.7.4197,PhysRevB.64.132410,Dutta2008} In addition, 
$T_{\mathrm{N}} \approx 148$~K was estimated a Curie-Weiss law analysis of the data obtained at 10 kOe.  
Ensemble, various analyses of the data consistently indicate $T_{\mathrm{N}} = 146 \pm 2$~K.

\subsection{ac susceptibility}

\begin{figure}[b]
\includegraphics[width=8.6cm]{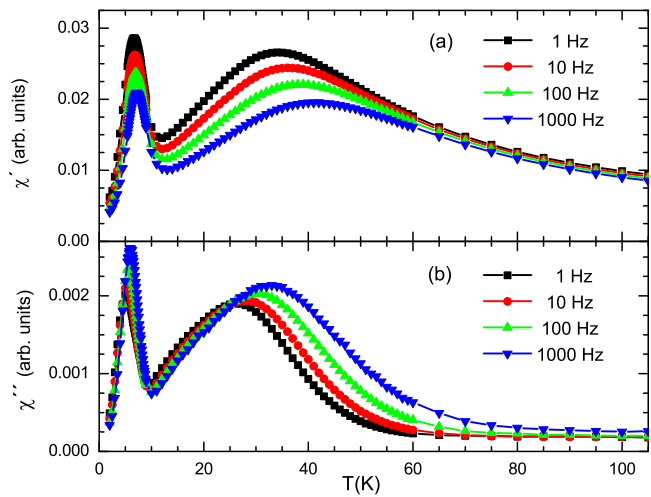}
\caption{\label{fig8}(Color online) Temperature dependence (a) of in-phase $\chi'(T)$ and (b) out-of-phase 
$\chi''(T)$ ac susceptibility at different frequencies of ac field. The data were taken at 1, 10, 100 and 1000 Hz 
as indicated in the figure. The first maximum is located at 6 K while the second maximum occurs at 37 K.}
\end{figure}

\begin{figure}[b]
\includegraphics[width=8.6cm]{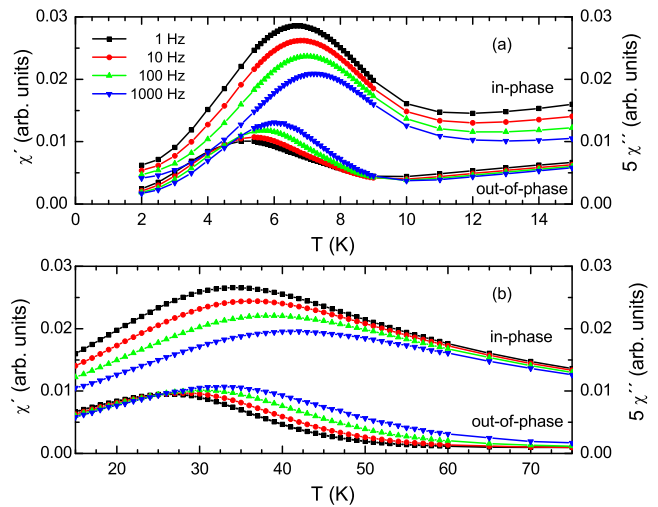}
\caption{\label{fig9}(Color online) Detail of the (a) low temperature (sharp) and (b) high temperature 
(broad) maxima in the temperature dependence of 
ac susceptibility. The in-phase $\chi'(T)$ and out-of-phase $\chi''(T)$ susceptibilities are plotted 
with the same scale in arbitrary units (arb. units) but are off-set for more pronounced comparison with 
the $\chi''(T)$ scale amplified by a factor of 5. }
\end{figure}

The real (in-phase), $\chi^{\prime}(T,\nu)$, and the imaginary (out-of-phase), $\chi^{\prime \prime}(T,\nu)$, 
of the magnetic susceptibility are shown in Figs.~\ref{fig8} and \ref{fig9} after ZFC.  The existence of 
two maxima are observed in both components near the temperatures of the peaks resolved in the dc study, 
Fig.~\ref{fig6} and Table~\ref{table1}, but the ac data reveal these feasures to be frequency dependent.
More specifically, the real part of the magnetic responses shifts toward higher temperatures for 
both maxima with increasing frequency, while the amplitudes of the magnetic signals decrease.  These 
trends are also reflected in the imaginary part of the magnetic responses although the amplitudes 
of the magnetic signals show a weak increase.  Generally speaking, these results for the dynamical 
response are characteristics of blocking or freezing processes, so additional inspection of the the 
thermal and frequency responses are necessary to clarify the nature of the peaks. 

The narrow low temperature peak located at 6 K exhibits the presence of a ``cusp-like'' maximum in 
$\chi'(T)$, which weakly depends on the frequency. Such characteristics are expected for ``ideal'' spin 
glasses.\cite{PhysRevB.60.12918,mydosh94}  Contrastingly, the high temperature peak of $\chi'(T)$ located 
near 37 K exhibits features associated with a broad distribution and possesses a strong frequency dependence.  
A useful and sensitive criterion to distinguishing between the freezing and the blocking processes 
lies in the determining the relative shift of the peak temperature in $\chi'$$(T)$ given 
by \cite{mydosh94}
\begin{equation}
p\;=\;\dfrac{\Delta T_{\mathrm{max}} }{T_{\mathrm{max}}\; \Delta\log \nu}\;\;\;,
\label{eq2}
\end{equation}
where $T_{\mathrm{max}}$ is the average value of the frequency dependent blocking/freezing temperature 
determined by the maximum of $\chi'(T)$, while $\Delta T_{\mathrm{max}}$ denotes the difference 
between $T_{\mathrm{max}}$  measured in the $\Delta\log \nu$  frequency interval. The parameter $p$ 
assumes values in the range $0.0045 - 0.06$ for atomic spin glasses\cite{PhysRevB.72.104433,mydosh94,Goya2004} 
and $0.10 - 0.13$ for non-interacting superparamagnets.\cite{PhysRevB.72.132409,PhysRevB.60.12918,mydosh94,C2CP22473A} 
These ranges can be compared to the values of $p$ obtained for low temperature and high temperature features observed 
in our CoO-Pt core-shell samples, where $p = 0.026$ calculated for low temperature maximum falls within the 
interval tpyically associated with a spin-glass state.  On the other hand, the value of $p = 0.08$ 
calculated for the high temperature maximum falls below the range usually associated with a non-interacting 
nanoparticle system,\cite{PhysRevB.72.132409,PhysRevB.60.12918,mydosh94,Goya2004} and thus the 
presence of inter-particle interactions can be inferred. Now that these qualitative associations have been 
made, quantitative analysis of the magnetic data will allow additional interpretations to be made.

\subsubsection{High temperature (broad) peak}
\begin{figure}[b]
\includegraphics[width=8.6cm]{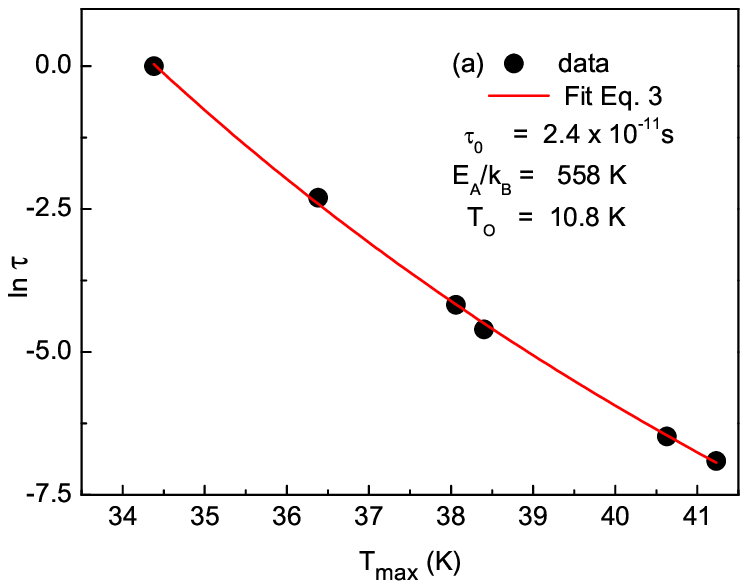}
\includegraphics[width=8.6cm]{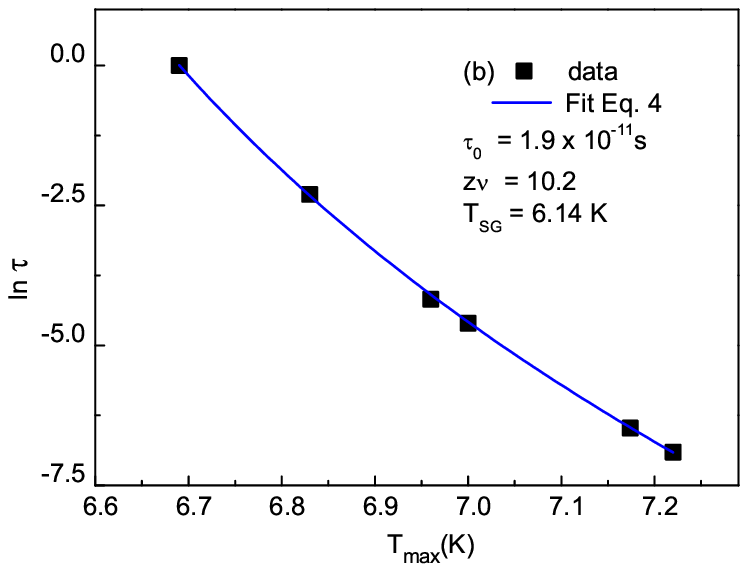}
\caption{\label{fig10}(Color online) Dependence of $\ln \tau$ vs. critical temperature $T_{\mathrm{f}}$ 
determined from the ac susceptibility at frequencies of 1, 10, 65, 100, 650 and 1000 Hz.
In (a), the high temperature (broad) peak data are fit by Eq.~3, 
while in (b), the low temperature (sharp) peak data are fit by Eq.~4., see text for details.}
\end{figure}

To quantitatively analyze the high temperature peak, the Stoner-Wohlfarth-Ne\'{e}l description can be 
used.\cite{Tadic200912,CVV2011}  In this description, an anisotropy energy barrier $E_{\mathrm{A}}$  
blocks the magnetic moments until a sufficient thermal activation energy $k_{\mathrm{B}}T$ relaxes spins 
from the blocked state to the superparamagnetic state. The dynamics of the non-interacting superparamagnets are described by
Ne\'{e}l-Arrhenius law,\cite{PhysRevB.60.12918,PhysRevB.72.104433} which can be written as  
\begin{equation}
\tau\;=\;\tau_{0} \,\exp \left( \dfrac{E_{\mathrm{A}}}{k_{\mathrm{B}}\, T_{\mathrm{max}}} \right)\;\;\;,
\label{eq3}
\end{equation}
where $\tau$ is the time associated with particle flips between two energy states, $\tau_{0}$ is an 
attempt frequency, and $T_{\mathrm{max}}$ is the temperature at which $\chi'(T)$ exhibits a 
maximum.\cite{Tadic200912,PhysRevB.72.104433,0295-5075-76-1-142} 
For non-interacting particles, typical values for $\tau_{0}$ are usually within the range 
$10^{-9} - 10^{-12}$~s. Fitting the experimental data to Eq.~(\ref{eq3}) yields $\tau_{0} = 6 \times 10^{-19}$~s, 
which is considerably lower than the values expected for non-interacting particles. So the breath of the 
analysis can be expanded by using the Vogel-Fulcher law,\cite{Tadic200912,PhysRevB.60.12918,0295-5075-76-1-142}
namely
\begin{equation}
\tau\;=\;\tau_{0}\, \exp \left( \dfrac{E_{\mathrm{A}}}{k_{\mathrm{B}}\, 
(T_{\mathrm{max}}-T _{0})}\right)\;\;\;,
\label{eq4}
\end{equation}
where $T_{0}$ accounts for a static interaction field due to the moments of surrounding 
particles.\cite{JAP2009}   
Fitting the data associated with the high temperature peak with Eq.~(\ref{eq4}) yields the reasonable results 
shown in Fig.~\ref{fig10}(a). The resulting values for the parameters are $\tau_{0}= 2.4 \times 10^{-11}$~s, 
$E_{\mathrm{A}}/k_{\mathrm{B}} = 558$~K and $T_{0} = 10.8 \pm 0.5$~K.  The anisotropy barrier $E_{A}$ is 
related with the uniaxial anisotropy constant $K_{A}$ trough relation $E_{A} = K_{A}V$, where 
$V$ is the volume of the nanoparticle.  For a particle with diameter $d = 4$~nm, 
$K_{A} = 2 \times 10^{6}$~erg/cm$^{3}$, which is a reasonable result.

\subsubsection{Low temperature (sharp) peak}
The temperature and frequency response of the ac susceptibility in the vicinity of the low temperature peak 
are dramaticaly different than the behavior observed near the high temperature peak.  Since the data are 
strikingly similar to observations reported in other 
nanosystems,\cite{PhysRevB.85.054429,CVV2011,PhysRevLett.79.5154} the analysis begins by differentiating 
between behavior indicative of superspin glass (SSG) and the atomic spin glass (SG) behavior.  
Specifically for spin glasses, the spin freezing temperature, $T_{\mathrm{f}} (\nu)$, can be defined as 
the temperature where the real part of ac susceptibility, $\chi'(T, \nu)$ manifests a 
maximum.\cite{PhysRevB.74.012411,Petracic2006192} Although $T_{\mathrm{f}}$ is often taken as a temperature 
at which $\chi'(T, \nu)$ is 0.98 times the equilibrium susceptibility, 
it is reasonable to define $T_{f} (\nu)$ as a temperature of maximum susceptibility in the $\chi'(T, \nu)$ 
curve for dynamical scaling analysis, as was demonstrated by  Gunnarsson \emph{et al.}\cite{PhysRevB.46.8227} 
and Djurberg \emph{et al.}\cite{PhysRevLett.79.5154} The dynamic scaling hypothesis provided 
that this system exhibits a conventional critical slowing down toward the transition temperature 
$T_{\mathrm{SG}}$, the variation of maximum relaxation time with transition temperature is described 
by\cite{PhysRevB.60.12918,PhysRevB.74.012411,PhysRevLett.79.5154,PhysRevB.67.174408} 
\begin{equation}
\tau\;=\;\tau_{0} \,\left(\dfrac{T_{\mathrm{\nu}}(\nu)-T_{\mathrm{SG}}}{T_{\mathrm{SG}}}\right)^{-z\upsilon}\;\;\;.
\label{eq5}
\end{equation}
Here, $\tau_0$ is the characteristic time scale for the spin dynamics, $T_{\mathrm{SG}}$ is the critical temperature 
for spin-glass ordering (this is equivalent to the $\nu \rightarrow 0$ value of $T_{f}$), $z\upsilon$ is a constant exponent, 
where $z$ is a dynamic exponent, and $\upsilon$ is the critical exponent characterizing the divergence of the correlation 
length $\xi$ given by\cite{PhysRevLett.79.5154}
\begin{equation}
\xi\;=\;\left( \dfrac{T_{\mathrm{SG}}}{(T_{\mathrm{SG}}-T_{\nu})}\right)^{\upsilon}\;\;\;.
\label{eq6}
\end{equation}
The agreement of experimental data with Eq.~(4) is shown in Fig.~\ref{fig10}(b), where the best fit yields the values of 
$z\upsilon = 10.2 \pm 0.6$, $T_{\mathrm{SG}} = 6.14 \pm 0.04$~K, and $\tau_{0} = 1.9 \times 10^{-11}$~s. 
These results are comparable with other atomic spin glass systems\cite{PhysRevB.64.174416,Morup2010,Zhang2003111,PhysRevB.82.094433} 
and nanoparticles superspin glass systems, where typical values for the parameters are $z\upsilon \sim  10$, 
$\tau_{0} \sim 10^{-11} - 10^{-13}$~s.  Ensemble, these results indicate the slow spin dynamics in the vicinity 
of the low temperarture peak.

\section{Summary}

The structure and the magnetic properties of fine nanoparticles composed of antiferromagnetic 
CoO cores coated by a Pt shell, prepared by a reverse micelle method, are presented. 
A suite of experimental probes were used to establish the structural and magnetic properties 
of the nanoparticles that possess a nominal diameter of 4~nm. Below about 6 K, the core-shell 
nanoparticles possess superspin glass properties.  At higher temperatures, up to about 37~K, 
the magnetism of the cores is blocked and interparticle interactions play an important role 
in the dynamical magnetic response.  This co-existence of blocking and freezing behaviors is 
consistent with the nanoparticles possessing an antiferromagnetically ordered core, 
with an uncompensated magnetic moment, and a magnetically disordered interlayer between CoO core and Pt shell.
Finally, due to their small diameters and the presence of the Pt shell, 
the cores experience magnetic ordering near 150~K.  Ultimately, these results provide 
benchmarks by which this system can be judged for potential use in microlectronic or 
biotechnology applications.

\begin{acknowledgments}
This work was supported by the Slovak Research and Development Agency under the contract APVV-0222-10 and 
APVV-0132-11 and VEGA projects of Ministry of Education of the Slovak Republic 
(No. 1/0583/11, No. 1/0861/12), by the Ministry of Education of the Czech Republic via 
ECOP program n. cz.1.07/2.3.00/30.0057 (\v{S}M), 
by the US National Science Foundation (NSF) via DMR-1202033 (MWM) and
DMR-1157490 (NHMFL), and by the Fulbright Commission of the Slovak Republic (MWM). The authors (AZ and VZ) 
would like to thank DESY/HASYLAB project No.~I-20110282 EC and also Dr.~N.~Murafa 
(AS CR, Rez, Czech Republic) for HRTEM measurements. Computing assistance from V.~Tk\'{a}\v{c} is also gratefully acknowledged.
\end{acknowledgments}

\bibliography{CoOPtvFeb27}

\end{document}